\documentclass[zpreprint,zbstpl]{zeus_paper}

\usepackage[english]{babel}

\chardef\usc=95
\chardef\til=126
\catcode`\@=11 
\DeclareRobustCommand\xdotspace{\futurelet\@let@token\@xdotspace}
\def\@xdotspace{%
  \ifx\@let@token.\else
  \ifx\@let@token\bgroup.\else
  \ifx\@let@token\egroup.\else
  \ifx\@let@token\/.\else
  \ifx\@let@token\ .\else
  \ifx\@let@token~.\else
  \ifx\@let@token!.\else
  \ifx\@let@token,.\else
  \ifx\@let@token:.\else
  \ifx\@let@token;.\else
  \ifx\@let@token?.\else
  \ifx\@let@token/.\else
  \ifx\@let@token'.\else
  \ifx\@let@token).\else
  \ifx\@let@token-.\else
  \ifx\@let@token\@xobeysp.\else
  \ifx\@let@token\space.\else
  \ifx\@let@token\@sptoken.\else
   .\space
   \fi\fi\fi\fi\fi\fi\fi\fi\fi\fi\fi\fi\fi\fi\fi\fi\fi\fi}
\catcode`\@=12 

\newcommand{\stru}[2]{%
   \relax\ifmmode\hbox{\vrule height#1 depth#2 width0pt}%
   \else\vrule height#1 depth#2 width0pt\fi}

\newcommand{\Ronum}[1]{\uppercase\expandafter{\romannumeral#1}}
\newcommand{\ronum}[1]{\expandafter{\romannumeral#1}}
\DeclareRobustCommand{\LaTeXZ}{%
  \LaTeX\kern-.05em4\kern-.1em
  {\raisebox{-0.2ex}{$\scriptstyle\text{ZEUS}$}}\xspace}

\DeclareMathAlphabet{\mathbf}{OT1}{cmr}{bx}{sl}
\newcommand{\eVdist}{\kern-0.06667em}

\newcommand{\gev}{{\,\text{Ge}\eVdist\text{V\/}}}

\newcommand{\slashfrac}[2]{%
  \raisebox{0.5ex}{\ensuremath #1}\kern-0.12em/\kern-0.08em
  \raisebox{-.8ex}{\ensuremath #2}}

\newcommand{\sqr}[3]{%
    {\vcenter{\hrule height.#3ex\hbox{\vrule width.#2ex height#1ex
     \kern#1ex\vrule width.#3ex}\hrule height.#2ex}}}

\catcode`\@=11 
\newcommand{\parenbar}{\mathpalette\p@renb@r}
\def\p@renb@r#1#2{\vbox{%
  \ifx#1\scriptscriptstyle \dimen@.7em\dimen@ii.2em\else
  \ifx#1\scriptstyle \dimen@.8em\dimen@ii.25em\else
  \dimen@1em\dimen@ii.4em\fi\fi \offinterlineskip
  \ialign{\hfill##\hfill\cr
    \vbox{\hrule width\dimen@ii}\cr
    \noalign{\vskip-.3ex}%
    \hbox to\dimen@{$\mathchar300\hfil\mathchar301$}\cr
    \noalign{\vskip-.3ex}%
    $#1#2$\cr}}}
\catcode`\@=12 

\newcommand{\IP}{{\rm I$\kern-0.01667em$P}\xspace}

\mathchardef\qsm=63
\mathchardef\pls=43
\mathchardef\mns=512
\mathchardef\plm=518
\mathchardef\eql=61
\mathchardef\smallleft=300
\mathchardef\smallright=301
\mathchardef\les=316
\mathchardef\gre=318
\mathchardef\leq=532
\mathchardef\grq=533

\catcode`\@=11 
\newcounter{pict@width}
\newcounter{pict@height}
\newlength{\pict@scale}
\newcommand{\psfigadd}[4]{%
\setcounter{pict@width}{1*\ratio{#2+\pict@scale/2}{\pict@scale}}
\setcounter{pict@height}{1*\ratio{#3+\pict@scale/2}{\pict@scale}}
\setlength{\unitlength}{\pict@scale}
\hbox to #2{\hspace{-\fill}\begin{picture}(\thepict@width,\thepict@height)
\put(0,0){\psfig{figure=#1,width=#2,height=#3,clip=}}
\SetScale{0.283466457}
\SetWidth{1.763889}
{#4}
\end{picture}}
}
\newcounter{pict@widthfst}
\newcounter{pict@widthscd}
\newcounter{pict@widthtot}
\newcommand{\psfigaddtwo}[7]{%
\setcounter{pict@widthfst}{1*\ratio{#2+\pict@scale/2}{\pict@scale}}
\setcounter{pict@widthscd}{1*\ratio{#2+#4+\pict@scale/2}{\pict@scale}}
\setcounter{pict@widthtot}{1*\ratio{#2+#4+#6+\pict@scale/2}{\pict@scale}}
\setcounter{pict@height}{1*\ratio{#3+\pict@scale/2}{\pict@scale}}
\setlength{\unitlength}{\pict@scale}
\hbox{\hspace{-\fill}\begin{picture}(\thepict@widthtot,\thepict@height)
\put(0,0){\psfig{figure=#1,width=#2,height=#3,clip=}}
\put(\thepict@widthscd,0){\psfig{figure=#5,width=#6,height=#3,clip=}}
\SetScale{0.283466457}
\SetWidth{1.763889}
{#7}
\end{picture}}
}
\newcommand{\psfigror}[4]{%
\setcounter{pict@width}{1*\ratio{#2+\pict@scale/2}{\pict@scale}}
\setcounter{pict@height}{1*\ratio{#3+\pict@scale/2}{\pict@scale}}
\setlength{\unitlength}{\pict@scale}
\hbox{\begin{picture}(\thepict@width,\thepict@height)
\put(0,\thepict@height){\psfig{figure=#1,width=#3,height=#2,clip=,angle=270}}
\SetScale{0.283466457}
\SetWidth{1.763889}
{#4}
\end{picture}}
}
\newcommand{\psfigrol}[4]{%
\setcounter{pict@width}{1*\ratio{#2+\pict@scale/2}{\pict@scale}}
\setcounter{pict@height}{1*\ratio{#3+\pict@scale/2}{\pict@scale}}
\setlength{\unitlength}{\pict@scale}
\hbox{\begin{picture}(\thepict@width,\thepict@height)
\put(0,0){\psfig{figure=#1,width=#3,height=#2,clip=,angle=90}}
\SetScale{0.283466457}
\SetWidth{1.763889}
{#4}
\end{picture}}
}
\catcode`\@=12 
\newlength\listtextwidth

\catcode`\@=11 
\newlength{\@tabfninsert}
\newlength{\@tabfnwidth}
\newcommand{\tabfootnote}[2]{%
  \setlength{\@tabfninsert}{0.8em}
  \setlength{\@tabfnwidth}{\textwidth}
  \addtolength{\@tabfnwidth}{-\@tabfninsert}
  \addtolength{\@tabfnwidth}{-0.4em}
  \noindent\makebox[\@tabfninsert][r]{\footnotesize$^{#1}$\hfil}\hfill%
  \parbox[t]{\@tabfnwidth}{\footnotesize #2\hfill}}
\catcode`\@=12 

\def\citeCTD{{\cite{%
nim:a279:290,*npps:b32:181,*nim:a338:254%
}}\xspace}

\def\citeCAL{{\cite{%
nim:a309:77,*nim:a309:101,*nim:a321:356,*nim:a336:23%
}}\xspace}

\begin{document}

\prepnum{{DESY--04--016}}

\title{Observation of isolated high-\mbox{$E_T$} photons in deep inelastic
scattering
}
                    
\author{ZEUS Collaboration}

\date{February 2004}

\abstract{
First measurements of cross sections for isolated prompt photon
production in deep
inelastic $ep$ scattering have been made using the ZEUS detector
at the HERA electron-proton collider using an 
integrated luminosity of 121 pb$^{-1}$.
 A signal for  
isolated photons in the transverse energy and rapidity ranges
$5 < E_T^{\gamma} < 10 \gev$ and  $-0.7 < \eta^{\gamma} < 0.9$
 was observed  for virtualities of
the exchanged photon of $Q^2 > 35 {\gev}{^2}$.
Cross sections are presented for inclusive prompt photons
and
for those accompanied by a single jet in the range $E_T^{\rm{jet}} \ge 6 
\gev$
and $-1.5 \le \eta^{\rm{jet}} < 1.8$.
Calculations at order
$ \alpha^3 \alpha_s $ describe the data reasonably well.}

\makezeustitle

\def\3{\ss}                                                                                        
\pagenumbering{Roman}                                                                              
\begin{center}                                                                                     
{                      \Large  The ZEUS Collaboration              }                               
\end{center}                                                                                       
  S.~Chekanov,                                                                                     
  M.~Derrick,                                                                                      
  D.~Krakauer,                                                                                     
  J.H.~Loizides$^{   1}$,                                                                          
  S.~Magill,                                                                                       
  S.~Miglioranzi$^{   1}$,                                                                         
  B.~Musgrave,                                                                                     
  J.~Repond,                                                                                       
  R.~Yoshida\\                                                                                     
 {\it Argonne National Laboratory, Argonne, Illinois 60439-4815}, USA~$^{n}$                       
\par \filbreak                                                                                     
  M.C.K.~Mattingly \\                                                                              
 {\it Andrews University, Berrien Springs, Michigan 49104-0380}, USA
% ...
\newpage                                                                                           
$^{\    1}$ also affiliated with University College London, London, UK \\                          
% ...
                                                           %
                                                           %
\newpage   
                                                           %
                                                           %
\begin{tabular}[h]{rp{14cm}}                                                                       
$^{a}$ &  supported by the Natural Sciences and Engineering Research                               
          Council of Canada (NSERC) \\                                                             
$^{b}$ &  supported by the German Federal Ministry for Education and                               
          Research (BMBF), under contract numbers HZ1GUA 2, HZ1GUB 0, HZ1PDA 5, HZ1VFA 5\\         
$^{c}$ &  supported by the MINERVA Gesellschaft f\"ur Forschung GmbH, the                          
          Israel Science Foundation, the U.S.-Israel Binational Science                            
          Foundation and the Benozyio Center                                                       
          for High Energy Physics\\                                                                
$^{d}$ &  supported by the German-Israeli Foundation and the Israel Science                        
          Foundation\\                                                                             
$^{e}$ &  supported by the Italian National Institute for Nuclear Physics (INFN) \\                
$^{f}$ &  supported by the Japanese Ministry of Education, Culture,                                
          Sports, Science and Technology (MEXT) and its grants for                                 
          Scientific Research\\                                                                    
$^{g}$ &  supported by the Korean Ministry of Education and Korea Science                          
          and Engineering Foundation\\                                                             
$^{h}$ &  supported by the Netherlands Foundation for Research on Matter (FOM)\\                   
$^{i}$ &  supported by the Polish State Committee for Scientific Research,                         
          grant no. 620/E-77/SPB/DESY/P-03/DZ 117/2003-2005\\                                      
$^{j}$ &  partially supported by the German Federal Ministry for Education                         
          and Research (BMBF)\\                                                                    
$^{k}$ &  partly supported by the Russian Ministry of Industry, Science                            
          and Technology through its grant for Scientific Research on High                         
          Energy Physics\\                                                                         
$^{l}$ &  supported by the Spanish Ministry of Education and Science                               
          through funds provided by CICYT\\                                                        
$^{m}$ &  supported by the Particle Physics and Astronomy Research Council, UK\\                   
$^{n}$ &  supported by the US Department of Energy\\                                               
$^{o}$ &  supported by the US National Science Foundation\\                                        
$^{p}$ &  supported by the Polish State Committee for Scientific Research,                         
          grant no. 112/E-356/SPUB/DESY/P-03/DZ 116/2003-2005,2 P03B 13922\\                       
$^{q}$ &  supported by the Polish State Committee for Scientific Research,                         
          grant no. 115/E-343/SPUB-M/DESY/P-03/DZ 121/2001-2002, 2 P03B 07022\\                    
\end{tabular}                                                                                      

\newpage
\pagenumbering{arabic} 
\pagestyle{plain}

\section{Introduction}
\label{sec-int}
Isolated photons in the final state with
high transverse momenta are a direct probe of the dynamics of hard
 subprocesses in high energy collisions, since these `prompt' photons are
 largely insensitive to the effects of hadronisation.
 Prompt photons have been studied in a number of hadronic experiments.
 Early evidence for such processes  came from the R806
experiment at the CERN ISR~\cite{zfp:c13:207}. More recently, the CDF and
D\O\ experiments at the Tevatron collider have performed a number of QCD 
tests
using prompt 
photons~\cite{prl:68:2734,pr:d48:2998,prl:73:2662,prl:70:2232,prl:77:5011,
prl:84:2786}.
In previous ZEUS publications, the production of prompt photons in 
photoproduction has been 
studied~\cite{pl:b413:201,pl:b472:175,pl:b511:19}.
  In the present letter, for the first time, prompt photon measurements in 
deep inelastic
scattering (DIS) are reported, both inclusively and accompanied by jets.
 These processes test QCD in a new way by studying processes containing
two different hard scales, $Q^2$,
the exchanged photon virtuality, and $E_T^{\gamma}$, 
the transverse energy
of the emitted prompt photon.

Prompt photons are produced in DIS at lowest order in QCD, as shown in 
Fig.
1. These
  processes have been calculated to order
 ${O} {(\alpha^3\alpha_s)}$
 by Gehrmann-DeRidder,
 Kramer and Spiesberger~\cite{np:b578:326}, including
 interference terms for
 initial- and final-state radiation from the electron.
 In contrast, leading-logarithm parton-shower Monte Carlo (MC) models do 
not
 naturally predict events with two hard scales.

In this letter, results are presented for the process 
$ep \rightarrow e\gamma X$, where $X$ is anything, and for
$ep \rightarrow e\gamma + {\rm{jet}} +Y$, where $Y$ does not contain 
further
jets within the acceptance of the measurement.
 Comparisons are made to MC predictions and
 also to  ${O} {(\alpha^3\alpha_s)}$
calculations for the photon-jet final state.

 \section{Experimental set-up and event selection} 

A data sample corresponding to an integrated luminosity of 121 pb$^{-1}$ 
was used, taken between 1996 and 2000. This sample is the sum of
38 pb$^{-1}$ of $e^+p$ data taken  
 at a centre-of-mass energy of $300 \gev$ and 68 pb$^{-1}$ taken at $318 
\gev$,
plus 16 pb$^{-1}$ of $e^-p$ data taken at $318 \gev$. A single set of 
results
is presented for this combined sample. The MC cross sections (see Section
3) differ by under $4\%$ at the two centre-of-mass energies,
 well within the precision of these measurements. 
 Differences between the cross-sections for $e^+p$ and $e^-p$
collisions are expected to be negligible~\cite{priv:kramer:2002}.

A description of the ZEUS detector is given
elsewhere~\cite{pl:b297:404}.  Of particular importance in the present
work
are the uranium calorimeter (CAL) and the central tracking detector
(CTD).  

The CAL~\citeCAL has an angular
coverage of 99.7\% of $4\pi$ and is
divided into three parts (FCAL, BCAL, RCAL), covering the 
angular ranges 
$2.6^{\circ}-36.7^{\circ}$, $36.7^{\circ}-129.1^{\circ}$ and 
$129.1^{\circ}-176.2^{\circ}$, respectively\footnote{ The
ZEUS coordinate system is a right-handed Cartesian system with the 
$Z$ axis pointing in the proton beam direction, refered to as 
the `forward direction', and the $X$ axis pointing left towards the 
centre
 of HERA. The coordinate origin is at the nominal interaction point.}.
Each part consists of towers longitudinally subdivided into
electromagnetic (EMC) and hadronic (HAC) cells.  The electromagnetic
section of the BCAL (BEMC) consists of cells of 23.3~cm length
azimuthally, representing 1/32 of the full $360^{\circ}$, and width of 
4.9~cm in the $Z$ direction at its inner face, at a  radius of 123.2~cm
 from the beam line.  These cells have a projective geometry as
viewed from the interaction point.  The profile of the electromagnetic
signals observed in clusters of cells in the BEMC
discriminates between those originating from photons or 
electrons\footnote{Hereafter `electron' refers both to electrons and
 positrons unless specified.}
and those originating from neutral-meson decays. The CAL energy
resolutions, as measured under test-beam conditions, are
$\sigma (E) /E = 0.18 /\sqrt{E}$ for electromagnetic showers and
$\sigma (E)/E = 0.35 /\sqrt{E}$ for hadrons, with $E$ in GeV.

 The CTD~\citeCTD is a cylindrical
drift chamber situated inside a superconducting solenoid. Using the
tracking information from the CTD, the vertex of an event can be
reconstructed with a resolution of 0.4 cm in $Z$ and 0.1 cm in $X,Y$.
In this analysis, the CTD tracks are used to reconstruct the event
vertex, and  are also used in the selection criteria for high-$E_T$ 
photons.

The luminosity was determined from the rate of the bremsstrahlung
process
$e p \rightarrow e \gamma p$, where the high-energy photon was
measured in a lead-scintillator 
calorimeter~\cite{desy-92-066,*zfp:c63:391,*acpp:b32:2025} located at 
$Z = -107{\rm m}$.

The DIS events were selected online using a trigger based on energy 
deposits in the CAL consistent with a scattered electron.
 Offline, events which passed DIS cuts
 similar to those used in previous analyses~\cite{epj:c21:443}
 were selected. In addition a photon candidate was required. 
The value of  $Q^2$, as reconstructed from the
 final-state electron, was required to be above $35~\rm{GeV}^2$.
The energy of the
  scattered electron was required to be above 10~GeV
 and its polar angle in the range $139.8^{\circ}$ to 
 $171.9^{\circ}$, in order to be well measured in the RCAL and well
 separated from the photon candidate.
  Events were required to have a reconstructed vertex
position within the range $|Z| <40$~cm and 
  $35 < \delta < 65 \gev$, where
$\delta = \sum_i E_i (1 - \cos \theta_i)$, $E_i$ is the energy of the
 $i$th CAL cell, $\theta_i$ is its polar angle and the sum runs over all
 cells.
 
 For the subset of events used in the photon-jet study, jets
  were reconstructed from CAL cells using a cone
 algorithm with radius  0.7 \cite{epj:c1:109} in the laboratory frame. 
Corrections for energy 
losses, principally due to uninstrumented material in front of
 the CAL, were evaluated using MC simulated events, and 
were typically +(10-15)\%
 for jets with measured energy above $6 \gev$~\cite{pl:b511:19}.

\section{Monte Carlo event simulation}

 The MC programs {\sc Pythia}~6.206~\cite{cpc:135:238} and
 {\sc Herwig}~6.1~\cite{cpc:67:465} were used to simulate prompt photon
emission for the study of event-reconstruction
 efficiency. 
 In both generators, the partonic
 processes are simulated using leading-order matrix elements, with
 the inclusion of initial- and final-state parton
 showers. Fragmentation into hadrons is performed using the
 Lund string model~\cite{prep:97:31}
 in the case of {\sc Pythia}, and a cluster
 model~\cite{np:b238:492} in the case of {\sc Herwig}. The events 
 generated using the {\sc Pythia} and {\sc Herwig} programs were used
 to correct for detector and
 acceptance effects. The corrections provided by {\sc Pythia} were used as
 default and those given by {\sc Herwig} were used to estimate the
 systematic uncertainties due to the treatment of the event
 dynamics and of parton showering and hadronisation. 
 The detector response to photons and  neutral mesons ($\pi^0$ and $\eta$)
 was simulated by using single-particle MC generated events.

 The generated events were passed through the ZEUS detector and trigger
 simulation programs based on
 {\sc Geant}~3.13~\cite{tech:cern-dd-ee-84-1}. They were reconstructed and
 analysed by the same programs as the data. The jet search was
 performed using the energy measured in the CAL cells in the
 same way as for the data. The same jet algorithm was also applied to
 the final-state particles.

To study the effects of electron radiation, 
simulations were made of deep inelastic scattering events
using the {\sc Heracles}~4.6.1~\cite{cpc:69:155-tmp-3cfb28c9, 
*spi:www:heracles}
 program with the
{\sc Djangoh}~\cite{spi:www:djangoh11} interface to
the MC generators that provide the hadronisation. 
The collinear radiative corrections were found to be small in the
kinematic region of this analysis and were neglected.

\section{Photon candidate selection}

 The identification of events containing an isolated prompt photon
 candidate
 follows closely the approach used in previous
analyses~\cite{pl:b413:201,pl:b472:175,pl:b511:19}.
 Events were
 selected on the basis of an isolated photon candidate detected in the 
BCAL. The algorithm selected predominantly electromagnetic clusters of
  cells within a small angular cone. Initially, 
 larger electromagnetic clusters than
 are typical of a single photon were accepted to estimate 
 backgrounds. Use of shower shapes as a discriminant,
as described
 below,  allowed subtraction of the backgrounds due to $\pi^0$ and
 $\eta$ production.

 It was required that the
 reconstructed transverse energy of the cluster satisfied
$E_{T}^{\gamma}>5 {\gev}$
 and the pseudorapidity satisfied
 $-0.7<\eta^{\gamma}<0.9$.
 The cut $E_{T}^{\gamma}< 10 {\gev}$ 
 was imposed to ensure that the $\pi^0$ and $\eta$ subtraction
 method was effective. 

 The photon candidate was well separated from the scattered electron.
 Monte Carlo simulations and  ${O} {(\alpha^3\alpha_s)}$
 calculations (see Section 6.2)  show that for electrons in the
 range defined in Section 2, most photons radiated from the electron
  fall outside the prompt-photon acceptance used in this analysis, though 
 they still give an important contribution to the cross section in the 
 kinematic region of the measurement.

 To reduce backgrounds, the photon-candidate cluster was required to
 be isolated by demanding  
  $\Delta r > 0.2$, where $\Delta r = \sqrt{\Delta \phi^2 + \Delta
 \eta^2}$,
 the distance to the nearest reconstructed track in $\eta -\phi$ space.
 It was further required that  $E_{T}^{\gamma} / E_{T}^{\rm{cone}} > 0.9$,
 where $E_T^{\rm{cone}}$ is the energy 
 within a cone in $\eta -\phi$ of radius 1.0 around
 the photon candidate.
 This energy isolation requirement suppresses the contribution from photon
 candidates produced within jets.  Deeply
 virtual Compton scattering (DVCS) events were removed by demanding at 
least two tracks reconstructed in the CTD, since in DVCS the final state
seen in the detector consists only of a photon and an electron which are  
well separated~\cite{pl:b517:47, zeusdvcs}.

The selected candidates were still dominated by
 neutral mesons, such as $\pi^{0}$ and $\eta$, which decay to photons.
 The single-photon signal 
 was statistically extracted from the background using BEMC
 energy-cluster
 shapes.  The first distribution considered was that of $\langle
\delta Z \rangle$, where
 $\langle \delta Z \rangle = \Sigma (E_{\rm{cell}} | Z_{\rm{cell}} -
\overline{Z} |)/
 \Sigma E_{\rm{cell}}$. 
 Here $E_{\rm{cell}}$ is the energy deposited in a BEMC cell,
 $Z_{\rm{cell}}$ is the cell number measured in the $Z$ direction and
$\overline{Z}$ is the energy-weighted mean of $Z_{\rm{cell}}$.
 Figure 2a  shows the $\langle \delta Z \rangle$
distribution for data, together with a fit based
 on photon shower shapes and a
simulation of single particles in the detector ($\pi^0$ and
 $\eta$). 
Clear peaks are visible at $\langle \delta Z \rangle \simeq 0.15$ due to 
single
photons and $\langle \delta Z \rangle \simeq 0.5$ due to 
$\pi^0 \rightarrow \gamma \gamma$, as well as a tail due to the decays of 
heavier particles to two or more photons.

 The photon shower shapes used were derived in two
ways: from DVCS data~\cite{zeusdvcs}, and from single-photon MC 
simulation.  In Fig. 2,  photons 
found in DVCS data events are shown. The results of the two shower-shape 
methods gave indistinguishable background subtractions and differed only 
by an overall scale
factor of $5\%$ on the acceptance of the prompt-photon signal.
 The DVCS method gave the higher acceptance, as the DVCS
 single-photon showers
 are slightly narrower than those from the MC showers. The MC method
 was used in this analysis, because of the higher
statistics available. This allows rapidity and energy dependences
of shower shapes to be modelled; a scale correction of  $5\%$ 
was then applied.
 
The $\eta$ contribution was determined from a fit to the $\langle \delta Z 
\rangle$ distribution above 0.65. After removing 
candidates with $\langle \delta Z \rangle > 0.65$,
the final background subtraction was performed using the variable
$f_{\rm{max}}$, defined as the ratio of the energy of the highest-energy
cell in an electromagnetic cluster to the total cluster 
energy.
When incident on the BEMC, single
 photons form narrow clusters, with most of the energy going into only
 one cell, giving an $f_{\rm{max}}$ distribution peaked close to unity.
 Because of the projective geometry of the BEMC, a photon entering 
at the boundary between two cells typically has $f_{\rm{max}} \simeq 
0.5$. Thus  the $f_{\rm{max}}$  distribution for single photons  
peaks close to 1.0 and extends down to 0.5.
 In contrast, the neutral mesons decay to more than one photon, forming
 larger clusters in the BEMC.
In each bin of a plotted physical quantity, events were divided into
 two classes, with high and low values of $f_{\rm{max}}$ respectively.
From the number of events in each class, as well as the ratios of the
corresponding numbers for the $f_{\rm{max}}$ distributions of the
 single-particle samples, the number of events in the given bin
was evaluated~\cite{pl:b413:201}. 

 A total of 1875 events with $\langle \delta Z \rangle < 0.65$ were 
selected,
 of which 877 have $f_{\rm{max}} > 0.75$, yielding a signal of
572 and a background of 1303 events. The fits and signal extraction 
procedure were repeated for each bin of each distribution.

 Studies based on single-particle MC samples showed that the 
 photon energy measured in the BCAL was on average less than the true
 value, owing to energy loss in the uninstrumented material in front of
 the BCAL. To compensate
 for this effect, a correction of typically 0.2 GeV was 
 added to the photon energy~\cite{pl:b511:19}.

\section{Systematic uncertainties}
The following sources of systematic uncertainty were investigated:
 variations of the nominal $f_{\rm{max}}$ spectra for the photon 
 affecting the signal extraction;  change in the
detector energy scale calibration by ${\pm 3 \%}$, reflecting the overall
energy scale uncertainty;
and a change in the energy cut in both MC events and data by ${\pm 10 \%}$
for photons. This last uncertainty is motivated 
by the r.m.s. differences between hadron-level generated and reconstructed
energies. Also included as a systematic uncertainty
 is the difference in estimated acceptance between {\sc Herwig} and 
{\sc Pythia}, which is
mostly well below the statistical uncertainty.
 A change of $\pm 20 \%$ in the hadronic energy cut for photon-jet events
 for both data and reconstructed Monte Carlo events,
 representing the r.m.s. difference between hadron-level and 
 reconstructed jet energies was considered as
 an additional systematic uncertainty.
 The uncertainty of 2.2\% on the luminosity measurement was 
 neglected in the differential cross sections but included in the
 total cross sections.

 The method used for background subtraction is more sensitive to
 the shape of the $f_{\rm{max}}$ distribution of the 
background than to that of the signal. The background shape is relatively
 insensitive to the $\pi^0 / \eta$ ratio and hence the results
 using DVCS and MC photons are very similar. 
 A study was made of the effect on the results of the fact that
 the fits fall below the data at high $\langle \delta Z \rangle$. 
This is due to
 events with large $E_T^\gamma$, where the contribution of events with 
more than
 one $\pi^0$ with a multi-$\pi^0$
 invariant mass above the $\eta$ mass is likely
 to be important. A fit was made to the high-$E_T^\gamma$ data excluding 
the
 region $\langle \delta Z \rangle > 1.0$. The change in the 
extracted signal was well below the statistical uncertainty in the bin.

\section{Results}
\subsection{Inclusive prompt photon production}

The cross section for inclusive prompt photon production,
$e p \rightarrow e \gamma X$, has been
measured  in the following kinematic region: $Q^2~>~35~\gev^2$,
$E_e~>~10~\gev$, $139.8^\circ < \theta_e < 171.8^\circ$, 
$-0.7~<~\eta^\gamma~<~0.9$ and $5~<~E_T^\gamma~<~10~\gev$, 
with photon isolation such that at least 90\% 
of the energy found in an $\eta - \phi$ cone of radius 1.0 around 
the photon is associated with the photon.
 The measured cross section is
\begin{center}
$\sigma (e p \rightarrow e \gamma X) = 5.64 \pm 0.58 \rm{(stat.)} 
^{+0.47}_{-0.72} \rm{(syst.)}~\rm{pb}$.
\end{center}
The predicted cross sections from {\sc Pythia} and {\sc Herwig}
are lower than the data by factors of approximately 2 and 8, respectively.
Figures 3a and 3b show the measured rapidity and transverse energy 
distributions, compared to MC predictions normalised to the data.
The data are also presented in Table 1.
Both {\sc Pythia} and {\sc Herwig} describe the $E_T^\gamma$ spectrum and
{\sc Herwig} describes the rapidity well.
Figure 3c shows the $Q^2$ distribution  of the data, again compared to 
MC predictions. The agreement of {\sc Pythia}
 with the data is reasonable, but  {\sc Herwig} fails to describe
 the measured $Q^2$ spectrum. 
 As discussed in Section 6.2, the  ${O} {(\alpha^3\alpha_s)}$
 calculations suggest that the 
 discrepancies between {\sc Pythia} and the data in the rate and photon 
 rapidity distribution may be due to the fact that wide-angle
 initial- and final-state radiation from the electron 
are not included in the MC calculations. 

\subsection{Prompt photon plus jet production}
 Owing to divergences in cross-section calculations for prompt
 photons, a comparison  to   ${O} {(\alpha^3\alpha_s)}$
QCD predictions in DIS can be made only when
 there is a jet accompanying an isolated prompt photon. Jets were 
reconstructed as described in Section 2.
 For events satisfying the criteria for isolated prompt photons described 
 above, jets were counted only if they had $E_T^{\rm{jet}}> 6 \gev$
 and $-1.5<\eta^{\rm{jet}}<1.8$.
The measured total cross section for photon plus a single
  jet within this kinematic region is
\begin{center}
$\sigma (e p \rightarrow e \gamma + {\rm{jet}}+Y) = 0.86 \pm 0.14
\rm{(stat.)} 
^{+0.44}_{-0.34} \rm{(syst.)}~\rm{pb}$.
\end{center}
  Figure 4  shows the differential cross sections for `prompt photon plus
 one jet' events, together with  MC predictions. 
The data are also presented in Table 2.
 The transverse
energies of the photon and the jet are well described by the MC 
calculations.
 {\sc Herwig} describes the 
photon rapidity  well but the jet pseudorapidity peaks at lower values.
 {\sc Pythia} describes the jet pseudorapidity well, but the photon rapidity
 peaks too far forward, as was also the case for inclusive photons.

Figure 5 shows the same data as Fig. 4, compared to the 
${O}(\alpha^3 \alpha_s)$
parton-level calculations of Kramer 
and Spiesberger~\cite{priv:kramer:2002}.
These include all possible initial- and final-state single photon and
 gluon radiation, together with appropriate 
vertex corrections, and their interference terms. Higher-order effects,
such as collinear bremsstrahlung in the same event as a hard non-collinear
photon, estimated to be a 4\% effect, are omitted.
These calculations use the phase-space-slicing method to cancel the
infrared and collinear singularities. The MRST parton 
distributions~\cite{epj:c4:463} were used
for the parametrisation of the proton structure. Parton-to-photon
 fragmentation functions were taken from Bourhis, Fontannaz and
 Guillet~\cite{epj:c2:529}.
 The  renormalisation scale was chosen to be the transverse energy of
the jet.
 The effect of changing this scale up or
 down by a factor of two, to estimate the possible contribution of
 unknown higher-order terms, is shown in Fig. 5.
 The predicted total cross section
 for the mixture of energies and beam charges used in this analysis is
 $1.33~\pm~0.07~\rm{pb}$, where the uncertainty corresponds to the change
 in the result when the renormalisation scale is varied by a factor of 
 two. This parton-level calculation is compatible with the data.

 By definition, the ${O}(\alpha^3 \alpha_s)$ 
 parton-level calculation does not include
 the effects of hadronisation.
 Hadronisation effects were investigated by comparing
 the parton-level {\sc Pythia} and {\sc Herwig} distributions with
  the hadron level. The effect of 
 hadronisation would be to reduce the predictions by 30\% to 40\%. 
 Because of the overall poor description of the data by the MC
 simulations, hadronisation corrections were not applied to the
  ${O}(\alpha^3 \alpha_s)$ calculation.

 The ${O}(\alpha^3 \alpha_s)$ calculation shows that 65\% of photons are
 emitted by the electron, concentrated at low photon rapidities, and the
 rest by quarks. The photon rapidity and jet pseudorapidity distributions
 for the latter component 
 resemble the {\sc Pythia} predictions, which include only such photons.
 Interference between these processes contributes
 only 2\% to the total. The transverse-energy distributions of the two
 processes are similar.
 The ${O}(\alpha^3 \alpha_s)$ calculation predicts a higher jet 
cross section at forward pseudorapidity and at low $E_T^\gamma$ than is 
seen in the data.
 
\section{Conclusions}
The first observation of prompt photon production in deep inelastic
scattering
 has been presented, together with distributions for accompanying jets.
 Leading-logarithm parton-shower Monte Carlo models
  for photon emission by quarks ({\sc Pythia} and
 {\sc Herwig}) are each able to describe some but not
 all of the features of the data. Both describe the transverse energy
 distribution well and {\sc Herwig} describes the photon rapidity well.
 Both models predict too low a cross section.

 The results have been compared to an 
 ${O} {(\alpha^3\alpha_s)}$ parton-level
 calculation for
 $e  p \rightarrow e \gamma + {\rm{jet}}+Y$ in the acceptance region of 
this 
measurement.  The level of agreement is
satisfactory in photon rapidity and jet transverse energy but only fair 
for photon
transverse energy and jet pseudorapidity.
 The total predicted cross section is consistent with the measured value. 

\newpage
\noindent {\Large\bf Acknowledgements}
\vspace{0.2cm}

We thank the DESY directorate for their strong support and encouragement.
The special efforts of the HERA group are gratefully acknowledged. We are
grateful for the support of the DESY computing and network services. The 
design, construction and installation of the ZEUS detector have been made 
possible by the ingenuity and effort of many people who are not listed as 
authors. We thank G.~Kramer and H.~Spiesberger for the 
 ${O} {(\alpha^3\alpha_s)}$ calculations.
 We also thank G.~Ingelman and T.~Sj\"ostrand for useful discussions.
 
\newpage
\clearpage
\providecommand{\etal}{et al.\xspace}
\providecommand{\coll}{Coll.\xspace}
\catcode`\@=11
\def\@bibitem#1{%
\ifmc@bstsupport
  \mc@iftail{#1}%
    {;\newline\ignorespaces}%
    {\ifmc@first\else.\fi\orig@bibitem{#1}}
  \mc@firstfalse
\else
  \mc@iftail{#1}%
    {\ignorespaces}%
    {\orig@bibitem{#1}}%
\fi}%
\catcode`\@=12

\newpage
\clearpage
\begin{table}[p]
\centering
 \begin{tabular}{|c|c|}\hline
  $\eta^\gamma$ &  ${\rm{d}}\sigma / {\rm{d}} \eta^\gamma$ (pb) \\
 \hline
  -0.7 to  -0.3 &  4.95 $\pm  0.78 ^{+0.51}_{-0.63}$  \\
  -0.3 to   0.1 &  5.20 $\pm  0.75 ^{+0.20}_{-0.42}$ \\
   0.1 to   0.5 &  2.12 $\pm  0.69 ^{+0.25}_{-0.38}$ \\
   0.5 to   0.9 &  1.82 $\pm  0.66 ^{+0.24}_{-0.32}$ \\
 \hline
       $E_T^\gamma$ (GeV) &  ${\rm{d}}\sigma / {\rm{d}}E_T^\gamma$ (pb
GeV$^{-1}$) \\
 \hline
  5.0 to  6.0 &  1.25 $\pm  0.40 ^{+0.05}_{-0.03}$ \\
  6.0 to  8.0 &  1.40 $\pm  0.20 ^{+0.06}_{-0.05}$ \\
  8.0 to  10.0 & 0.79 $\pm  0.10 ^{+0.10}_{-0.01}$ \\
 \hline
 \end{tabular}
\vspace{1cm}
\caption{\small Differential cross sections for inclusive
 production of isolated photons with $-0.7 < \eta^\gamma < 0.9$,
 for $5 < E_T^\gamma < 10 \gev$.
The first uncertainty is statistical, the second is systematic. 
\protect\\[00mm] }
\label{tabsec1}
\end{table}

\begin{table}[p]
\centering
 \begin{tabular}{|c|c|c|c|}\hline

  $\eta^\gamma$ & ${\rm{d}}\sigma / {\rm{d}} \eta^\gamma$ (pb) &
  $\eta^{\rm{jet}}$ & ${\rm{d}}\sigma / {\rm{d}} \eta^{\rm{jet}}$
(pb) \\
 \hline
  -0.7 to  -0.3 &  0.81 $\pm  0.20 ^{+0.39}_{-0.27}$ &
  -1.5 to  -0.84&  0.087 $\pm 0.043 ^{+0.059}_{-0.018}$  \\
  -0.3 to  0.1 &  0.77 $\pm  0.19 ^{+0.33}_{-0.21}$ &
  -0.84 to  -0.18 &0.118 $\pm 0.068 ^{+0.20}_{-0.10}$  \\
  0.1 to  0.5 &  0.16 $\pm  0.16 ^{+0.18}_{-0.17}$ &
  -0.18 to  0.48 & 0.47 $\pm 0.13^{+0.25}_{-0.10}$  \\
  0.5 to  0.9 &   0.30 $\pm  0.15 ^{+0.19}_{-0.15}$ &
  0.48 to  1.14 &  0.24 $\pm 0.10 ^{+0.10}_{-0.06}$ \\
   &   &
   1.14 to 1.8 &   0.41 $\pm 0.12 ^{+0.10}_{-0.18}$ \\
 \hline
       $E_T^\gamma$ (GeV) &  ${\rm{d}}\sigma / {\rm{d}}E_T^\gamma$ (pb
GeV$^{-1}$) &
       $E_T^{\rm{jet}}$ (GeV) &  ${\rm{d}}\sigma /
      {\rm{d}}E_T^{\rm{jet}}$ (pb GeV$^{-1}$) \\
 \hline
  5.0 to  6.0 &  0.136 $\pm  0.090 ^{+0.060}_{-0.043}$ &
  6.0 to  8.0 &  0.177 $\pm  0.044 ^{+0.016}_{-0.028}$  \\
  6.0 to  8.0 &  0.258 $\pm  0.050 ^{+0.069}_{-0.041}$ &
  8.0 to  10.0 & 0.132 $\pm  0.049 ^{+0.057}_{-0.022}$ \\
  8.0 to  10.0 &  0.107 $\pm 0.029 ^{+0.046}_{-0.001}$ &
  10.0 to  12.0 & 0.082 $\pm 0.032 ^{+0.068}_{-0.018}$  \\
 & &  12.0 to  16.0 & 0.046 $\pm 0.031 ^{+0.023}_{-0.021}$ \\
 \hline
 \end{tabular}
\vspace{1cm}
\caption{\small Differential cross sections for
 production of isolated photons plus one jet with $-0.7 < \eta^\gamma
 < 0.9$, for $5 < E_T^\gamma < 10 \gev$ and $-1.5 < \eta^{\rm{jet}}
 < 1.8$, for $E_T^{\rm{jet}} > 6 \gev$.
The first uncertainty is statistical, the second is
 systematic.
\protect\\[00mm]} 
\label{tabsec2}
\end{table}

\newpage
\clearpage
\begin{figure}[p]
\vfill
\begin{center}
\epsfig{file=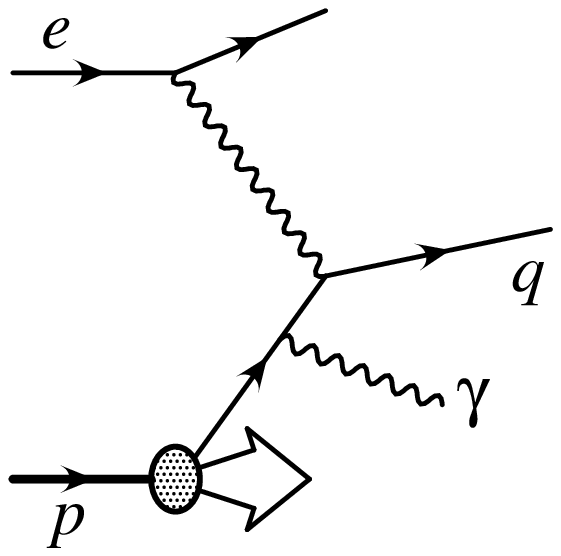,width=3cm}
\epsfig{file=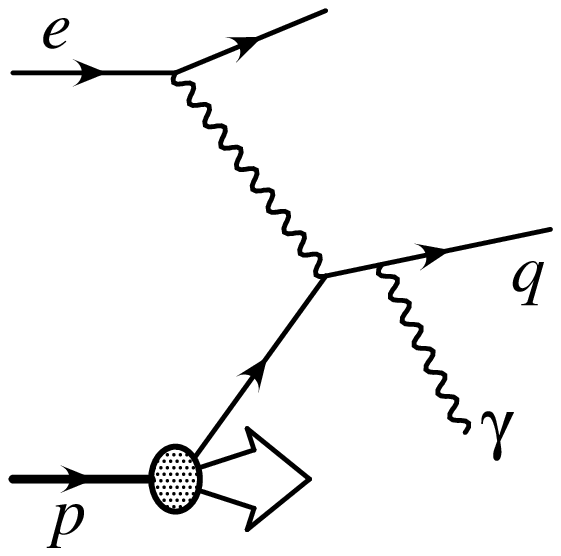,width=3cm}
\epsfig{file=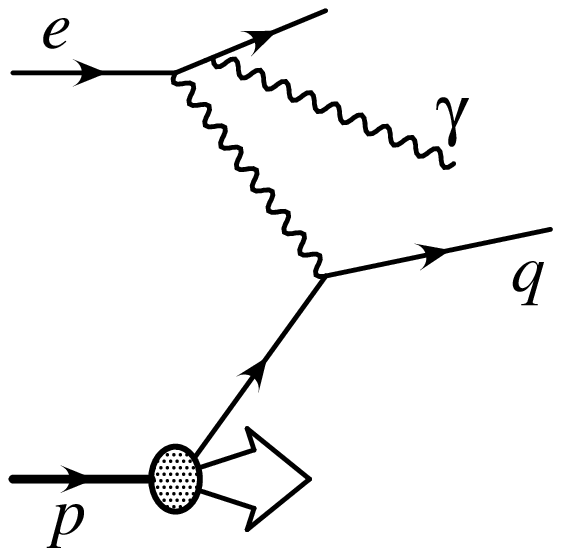,width=3cm}
\epsfig{file=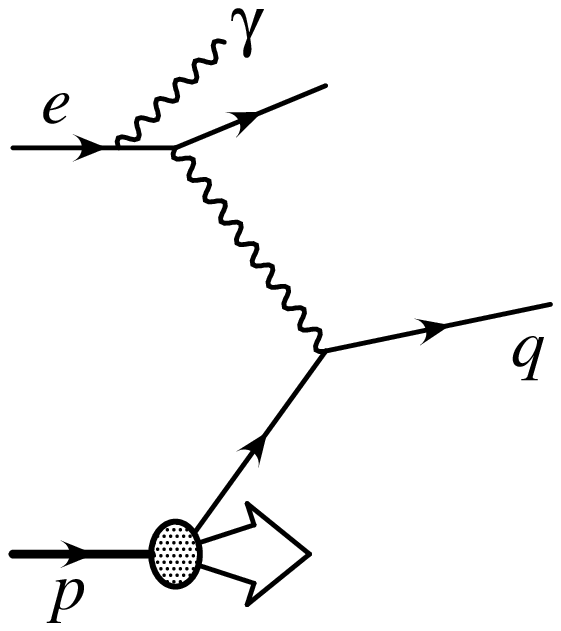,width=3cm}
\end{center}
\caption{\small The lowest-order tree-level diagrams for prompt photon 
production in 
$ep$ scattering. Vertex corrections enter at the same order.}
\label{fig1}
\vfill
\end{figure}

\newpage
\clearpage
\begin{figure}[p]
\vfill
\setlength{\unitlength}{1.0cm}
\begin{picture} (18.0,18.0)
\put (4.0,9.0){\epsfig{figure=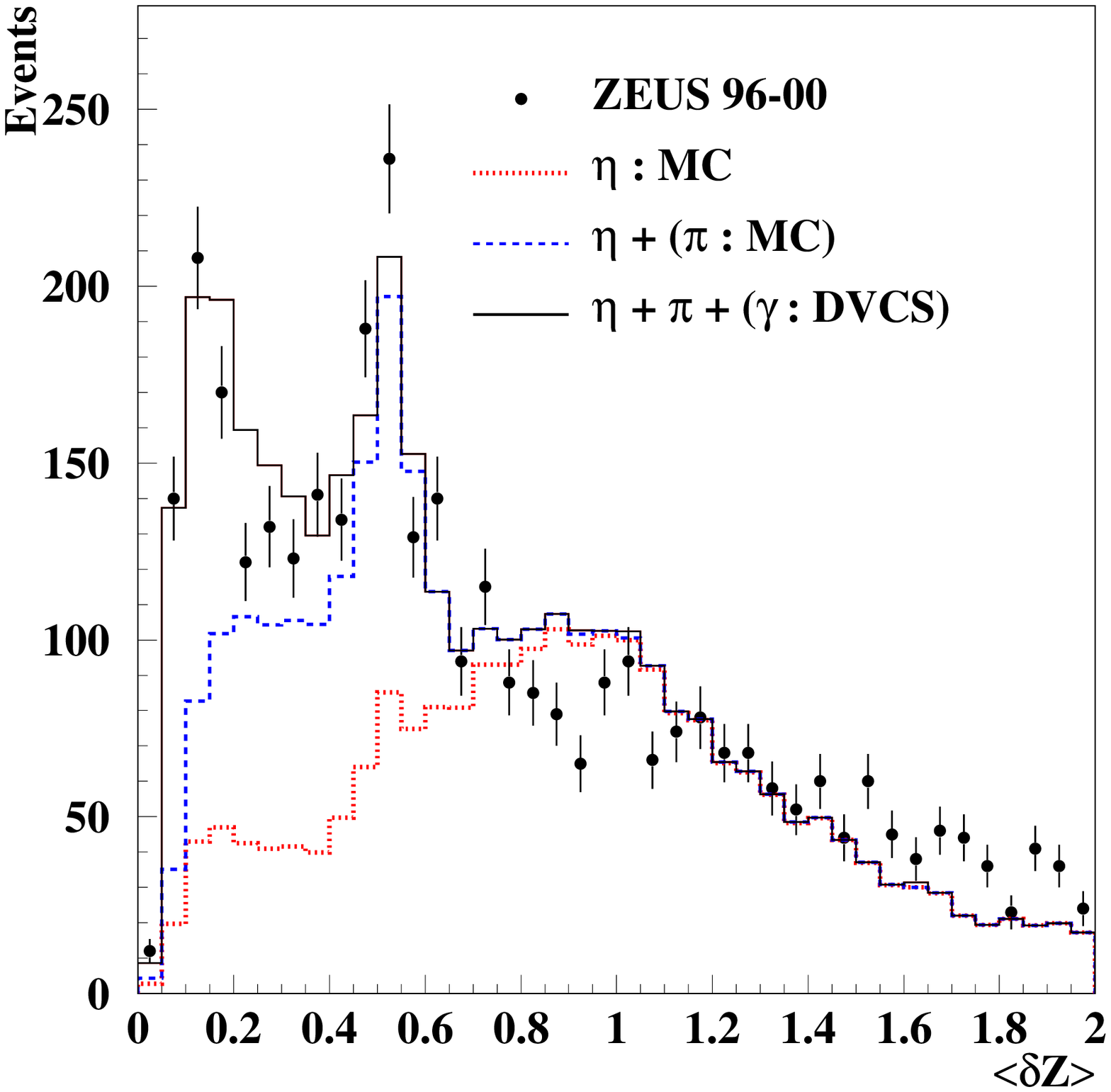,width=10cm}}
\put (4.0,0.0){\epsfig{figure=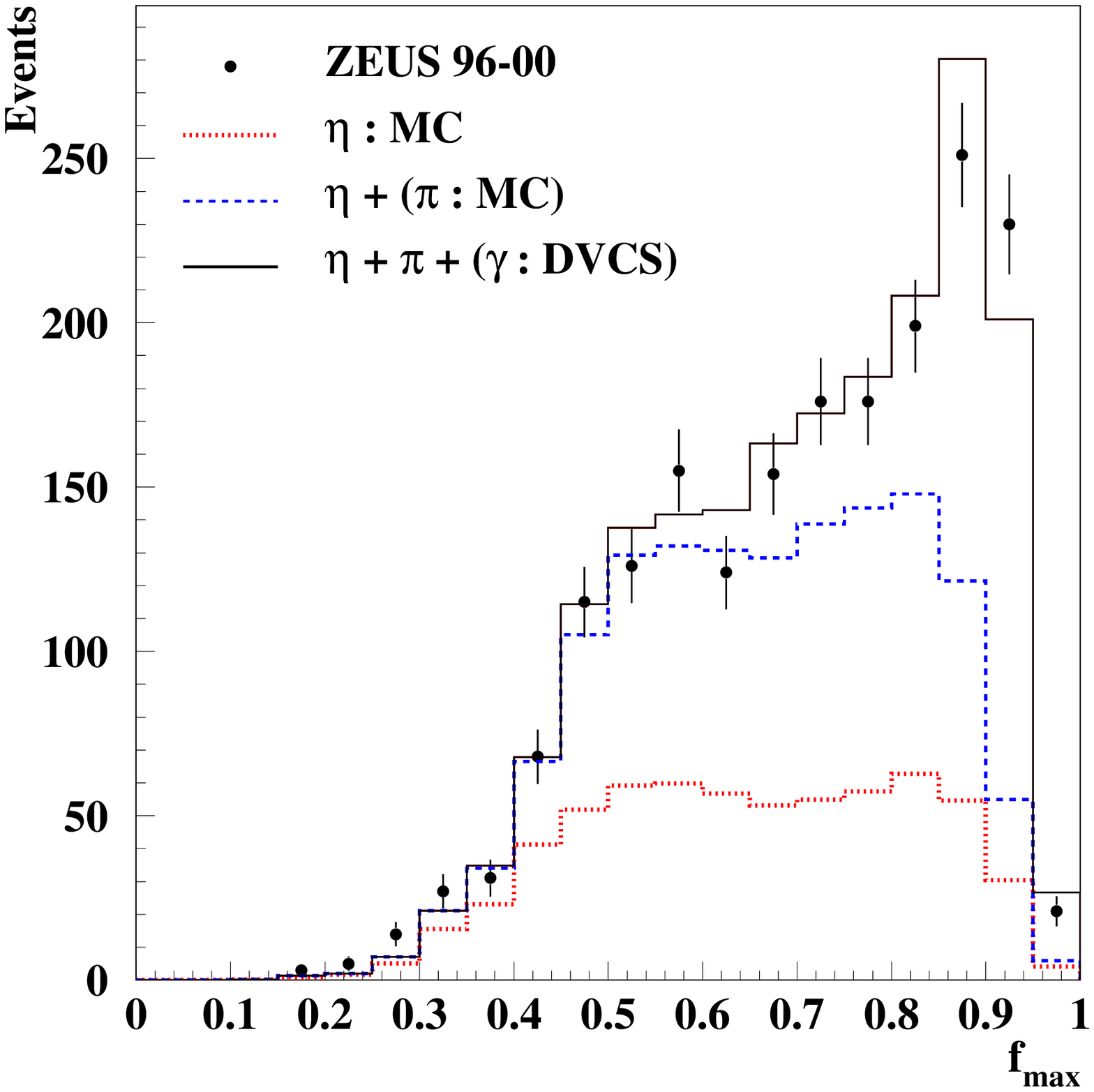,width=10cm}}
\put(0.9,18.5){\bf\Large\centerline{ZEUS}}
\put (12.3,17.5){\bf\small (a)}
\put (12.3,8.5){\bf\small (b)}
\end{picture}
\caption{\small (a) Distribution of $\langle \delta Z \rangle$ for 
prompt photon candidates in
selected events. (b) Distribution of $f_{\rm{max}}$ after a cut on
$\langle \delta Z \rangle <0.65$. Also given are fitted distributions for 
Monte Carlo $\eta$ mesons, $\pi^0 + \eta$ and $\pi^0 + \eta + 
\gamma$ (where the $\gamma$ is taken from DVCS data),  with similar 
selection criteria and $E_T^\gamma$ spectrum to the observed candidates.}
\label{fig2}
\vfill
\end{figure}

\newpage
\clearpage
\begin{figure}[p]
\setlength{\unitlength}{1.0cm}
\begin{picture}(18.0,18.0)
\put(0.0,10.0){\epsfig{figure=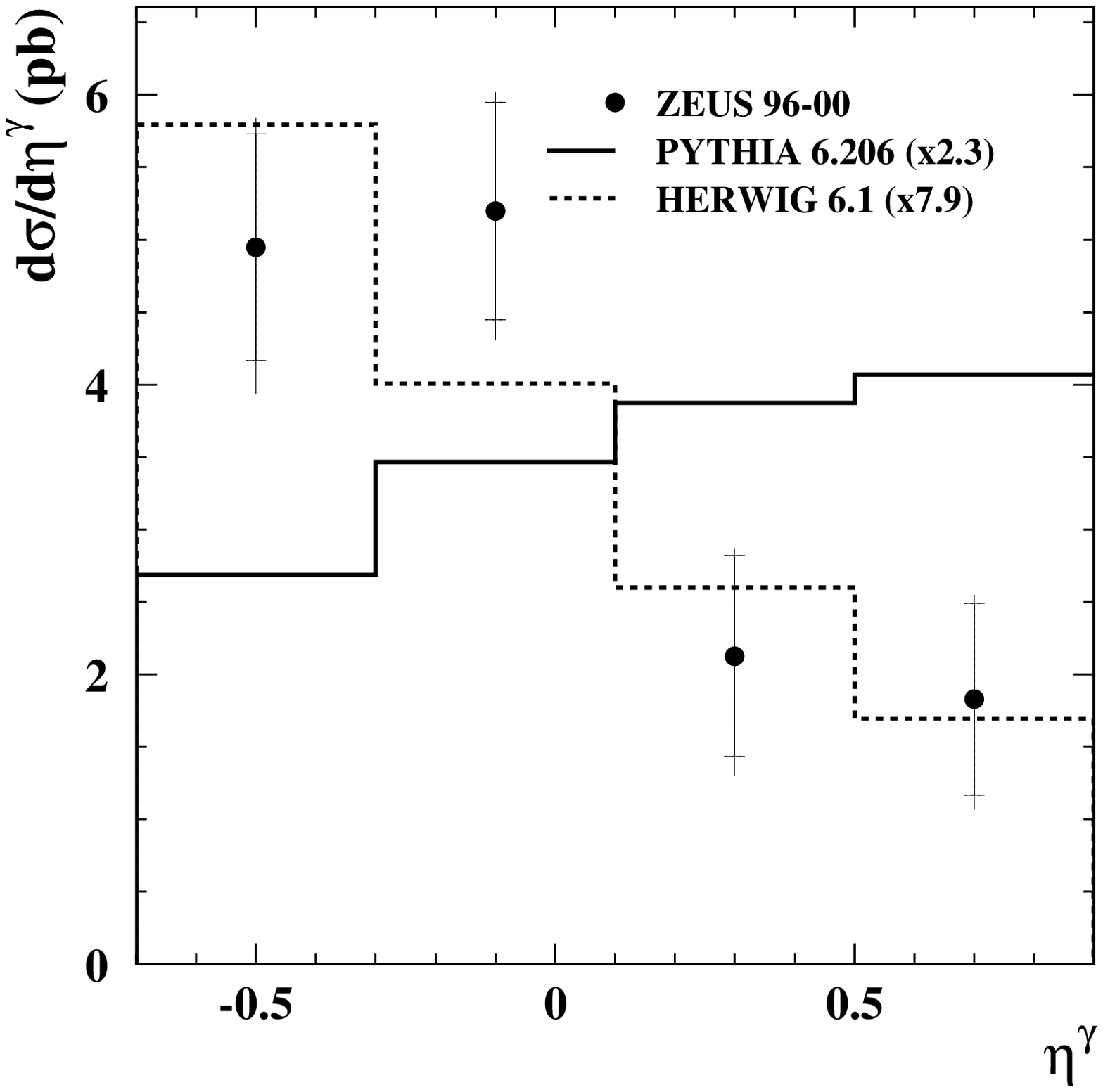,width=9cm}} 
\put(9.0,10.0){\epsfig{figure=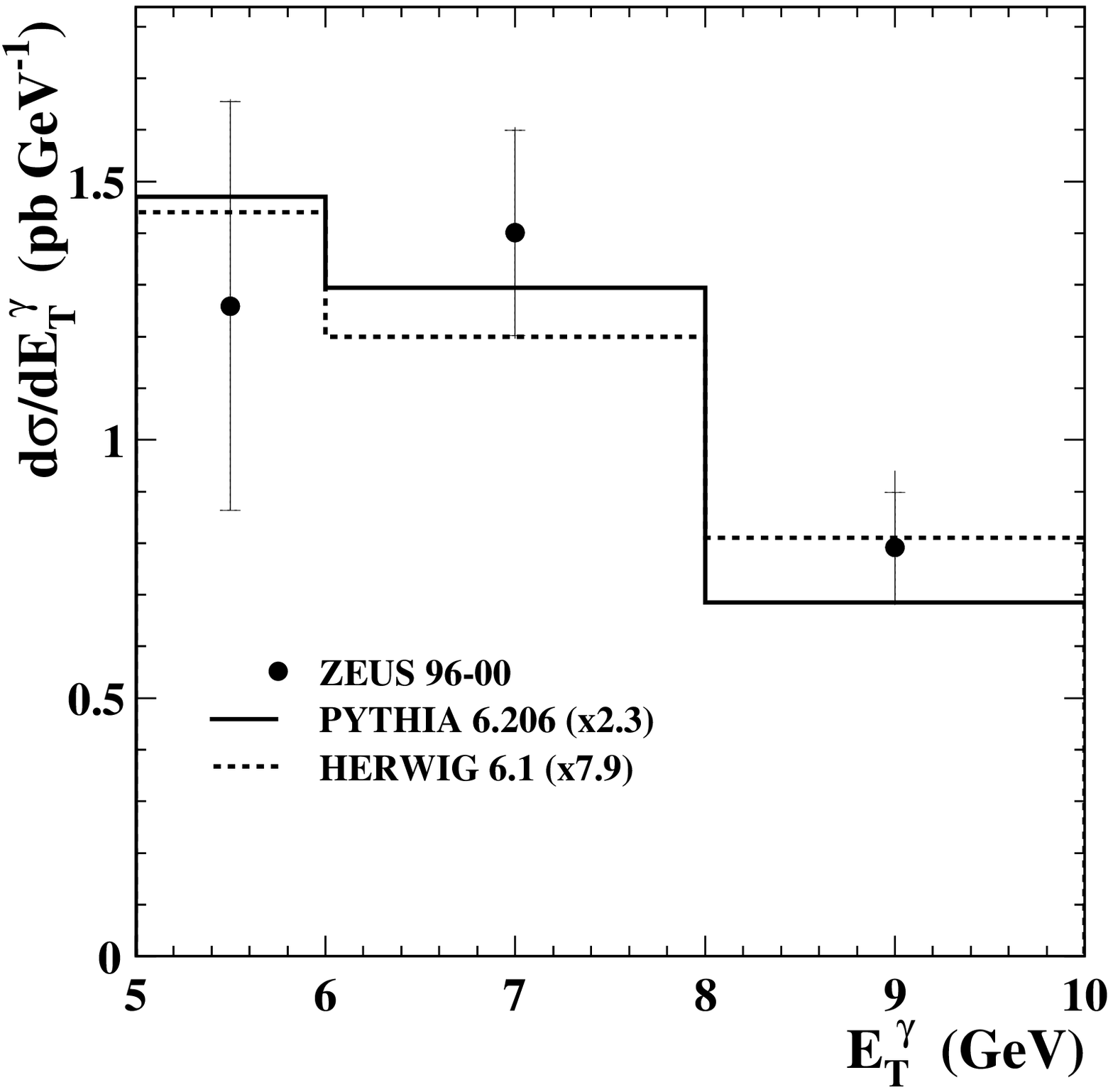,width=9cm}}
\put(0.0,0.6){\epsfig{file=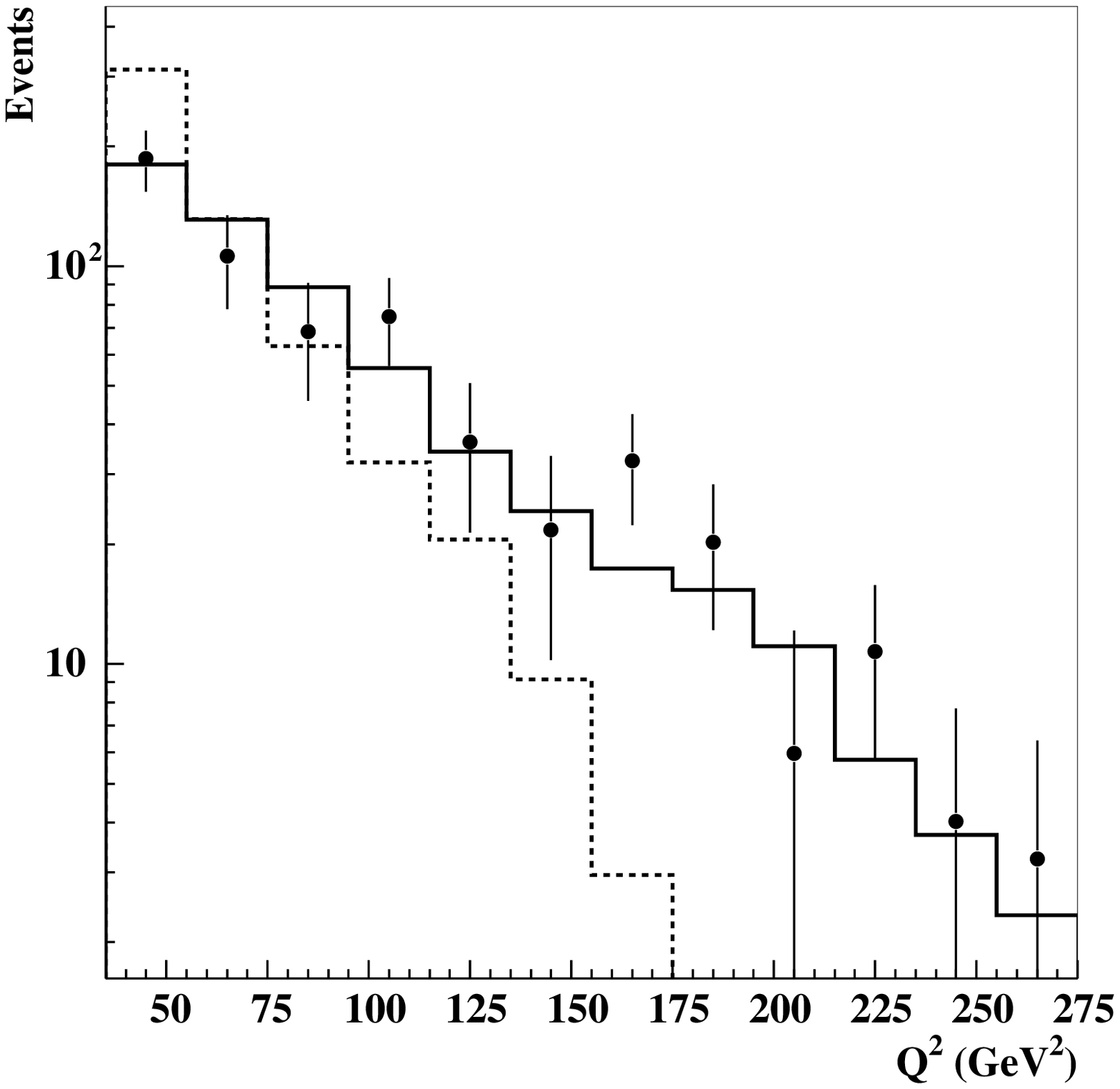,width=9cm}}
\put(-0.1,0.5){\epsfig{file=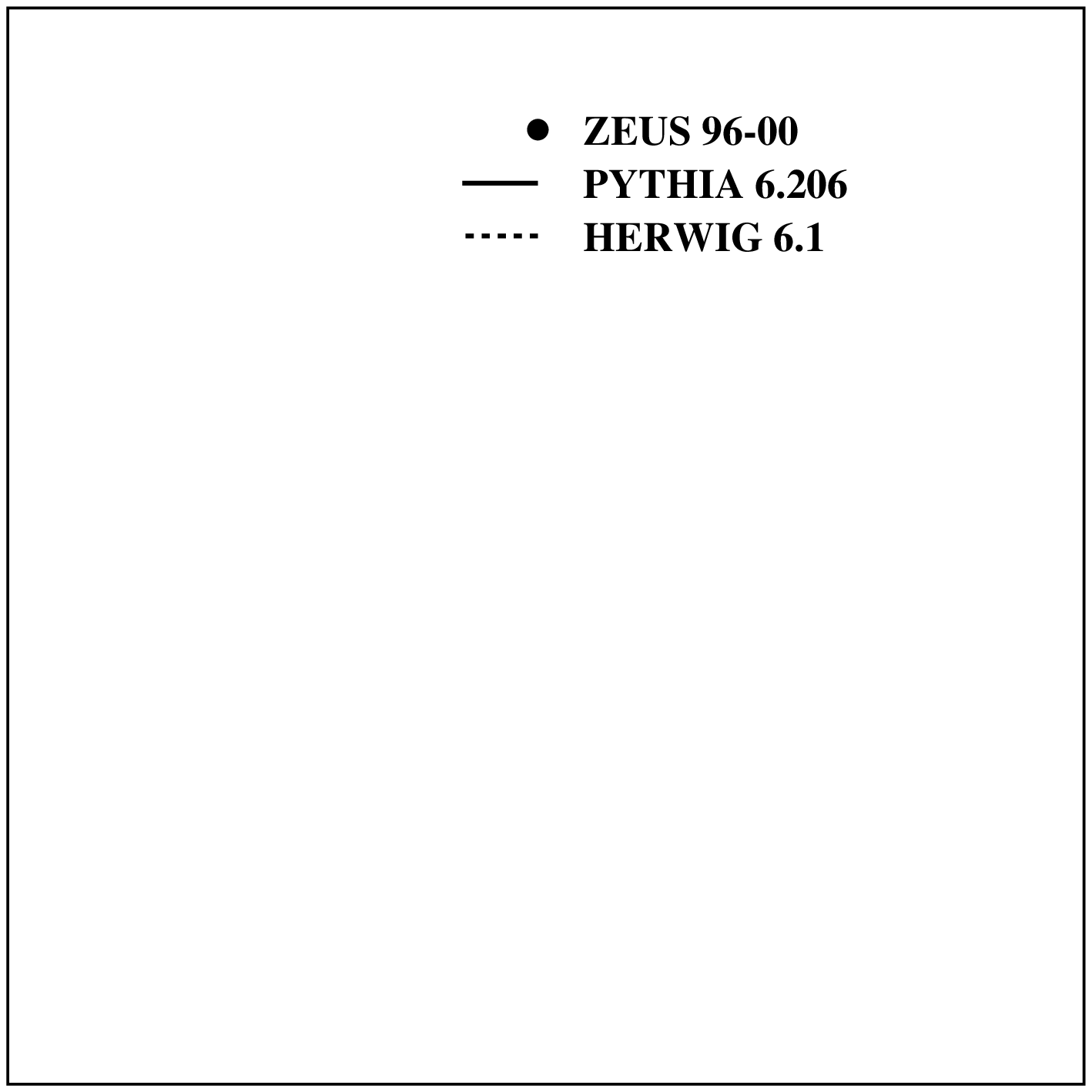,width=9cm}}
\put(0.9,18.5){\bf\Large\centerline{ZEUS}}
\put(7.3,17.4){\bf\small (a)}
\put(16.2,17.4){\bf\small (b)}
\put(7.3,8.0){\bf\small (c)}
\end{picture}
\caption{\small Inclusive prompt-photon differential cross section
(a) in rapidity,
 (b) in transverse energy, in the range
 $-0.7 < \eta^\gamma < 0.9$  and $5 < E_T^\gamma < 10 {\rm\gev}$. The 
inner error bars 
are statistical while the outer represent systematic uncertainties added
in quadrature. 
 (c) Distribution of $Q^2$. In each case
the histograms show MC predictions, normalised to data.}
\label{fig3}
\end{figure}

        \newpage
\clearpage
\begin{figure}[p]
\vfill
\setlength{\unitlength}{1.0cm}
\begin{picture} (18.0,18.0)
\put (0.0,10.0){\epsfig{figure=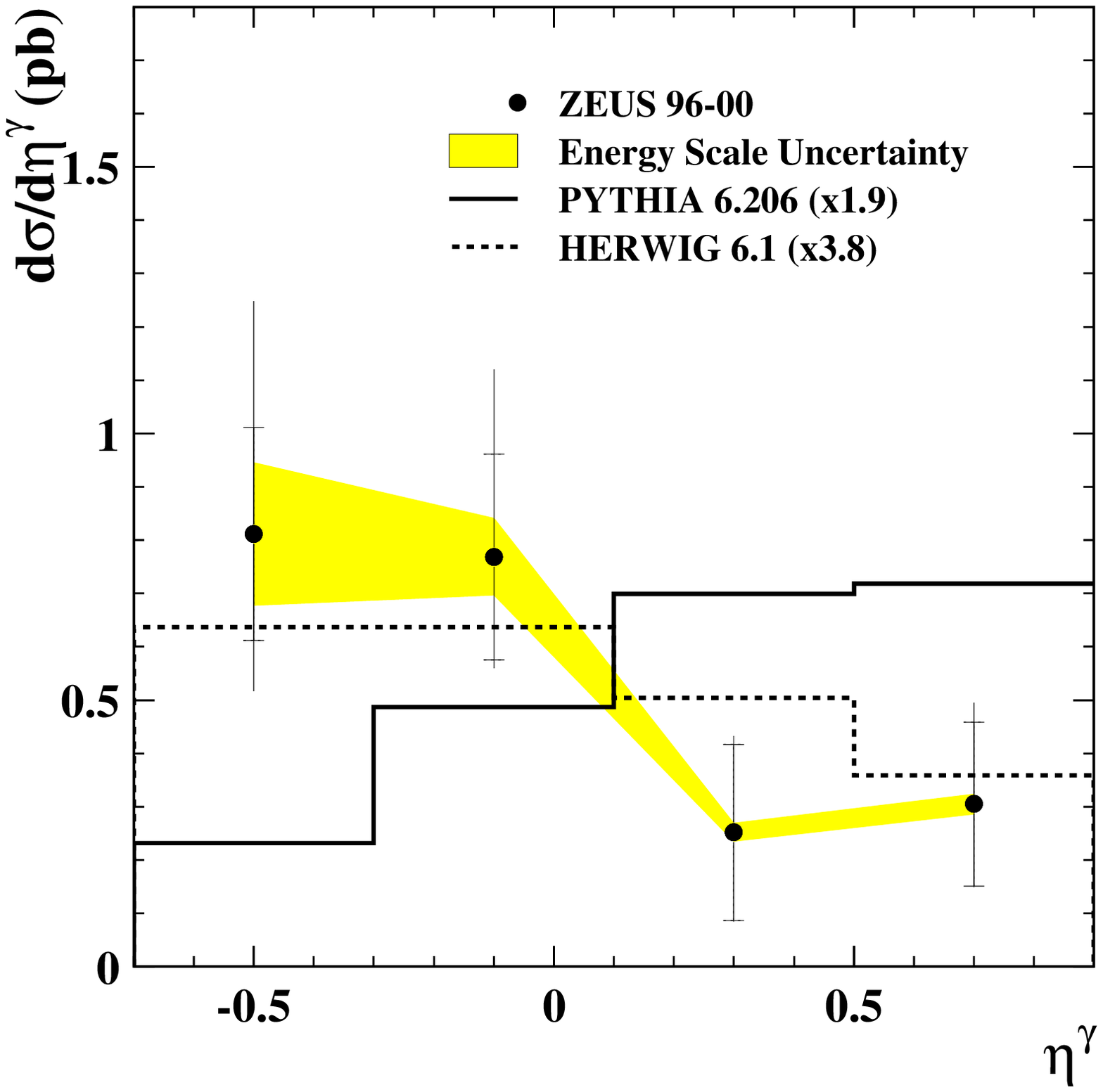,width=9cm}}
\put (9.0,10.0){\epsfig{figure=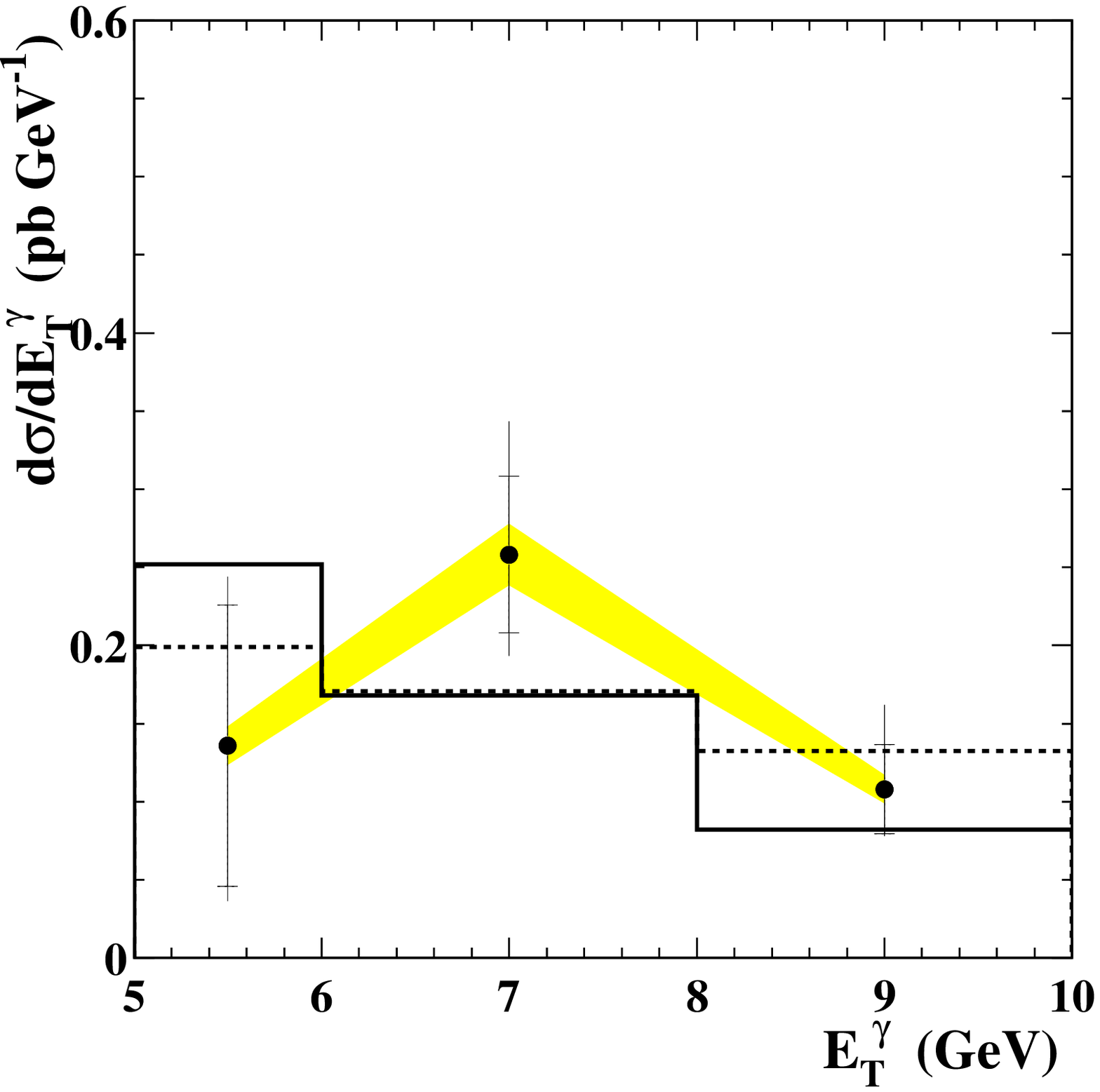,width=9cm}}
\put (0.0,0.6){\epsfig{figure=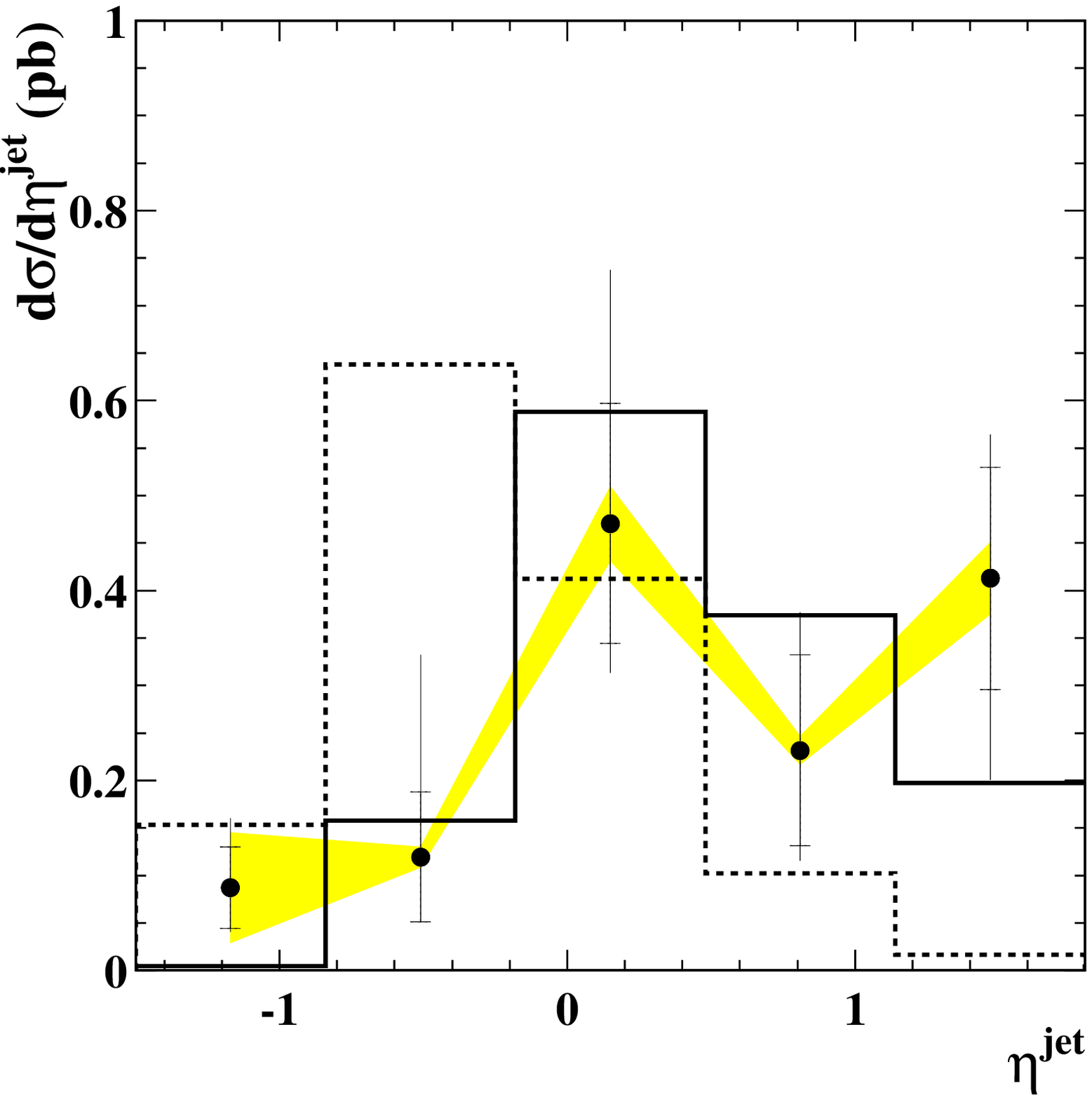,width=9cm}}
\put (9.0,0.6){\epsfig{figure=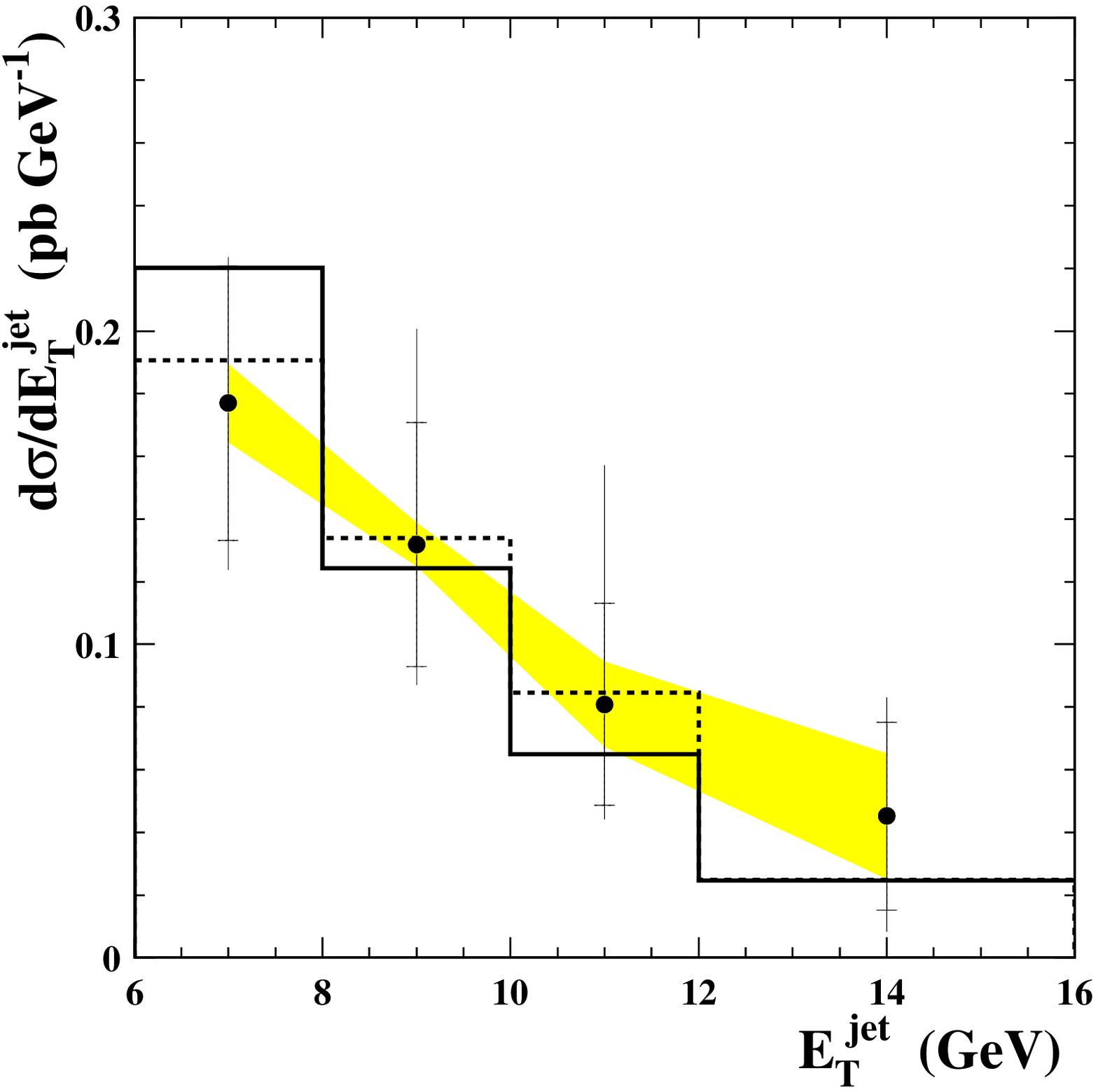,width=9cm}}
\put(0.9,18.5){\bf\Large\centerline{ZEUS}}
\put (7.3,17.4){\bf\small (a)}
\put (16.2,17.4){\bf\small (b)}
\put (7.3,8.0){\bf\small (c)}
\put (16.2,8.0){\bf\small (d)}
\end{picture}
\caption{\small Cross section for prompt-photon-plus-jet production
differential in
(a) photon rapidity,
(b) photon transverse energy,
(c) jet pseudorapidity,
 (d) jet transverse energy,
 for events with a photon in the range
$-0.7 < \eta^\gamma < 0.9$ and $5 < E_T^\gamma < 10 {\rm\gev}$ and
one jet in the range $-1.5 < \eta^{\rm{jet}} < 1.8$ and
$E_T^{\rm{jet}} > 6 {\rm\gev}$. The inner error bars are statistical and 
the outer represent systematic uncertainties added in quadrature.
The band around the data points
shows the effect of calorimeter energy-scale uncertainty.
 The histograms show Monte Carlo predictions, normalised to the
 data.}

\label{fig4}
\vfill
\end{figure}

\newpage
\clearpage
\begin{figure}[p]
\vfill
\setlength{\unitlength}{1.0cm}
\begin{picture} (18.0,18.0)
\put (0.0,10.0){\epsfig{figure=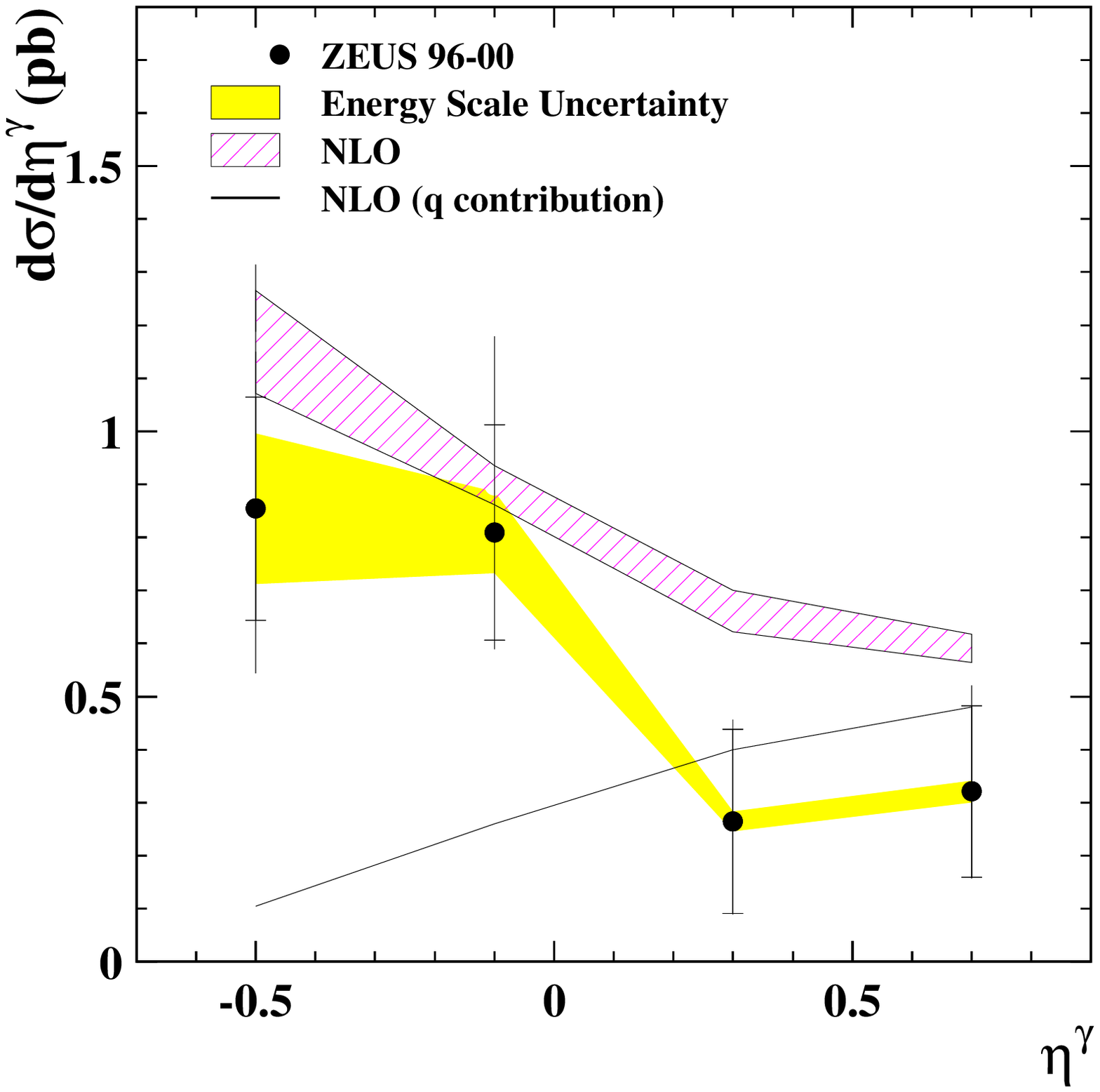,width=9cm}}
\put (9.0,10.0){\epsfig{figure=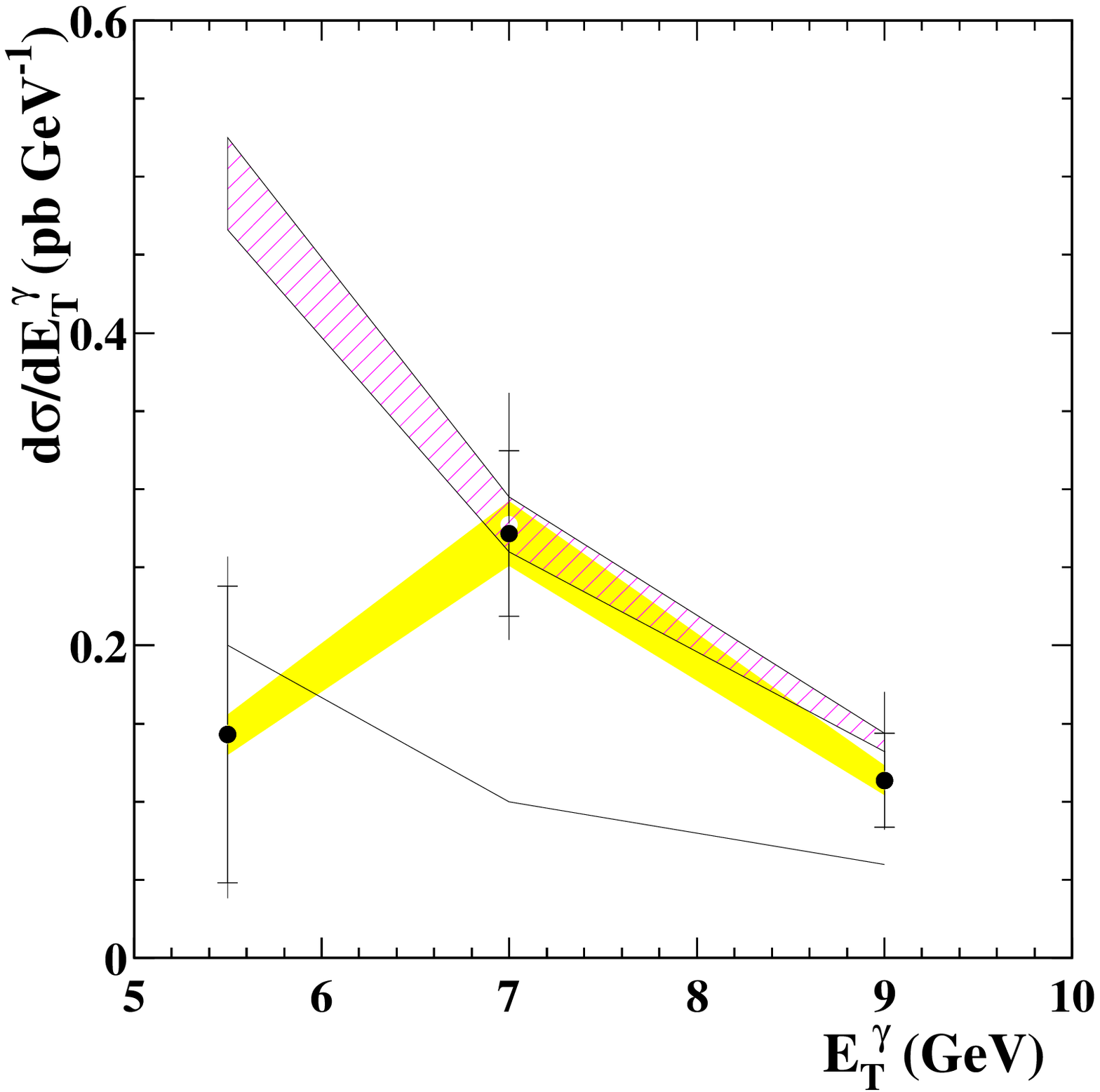,width=9cm}}
\put (0.0,0.6){\epsfig{figure=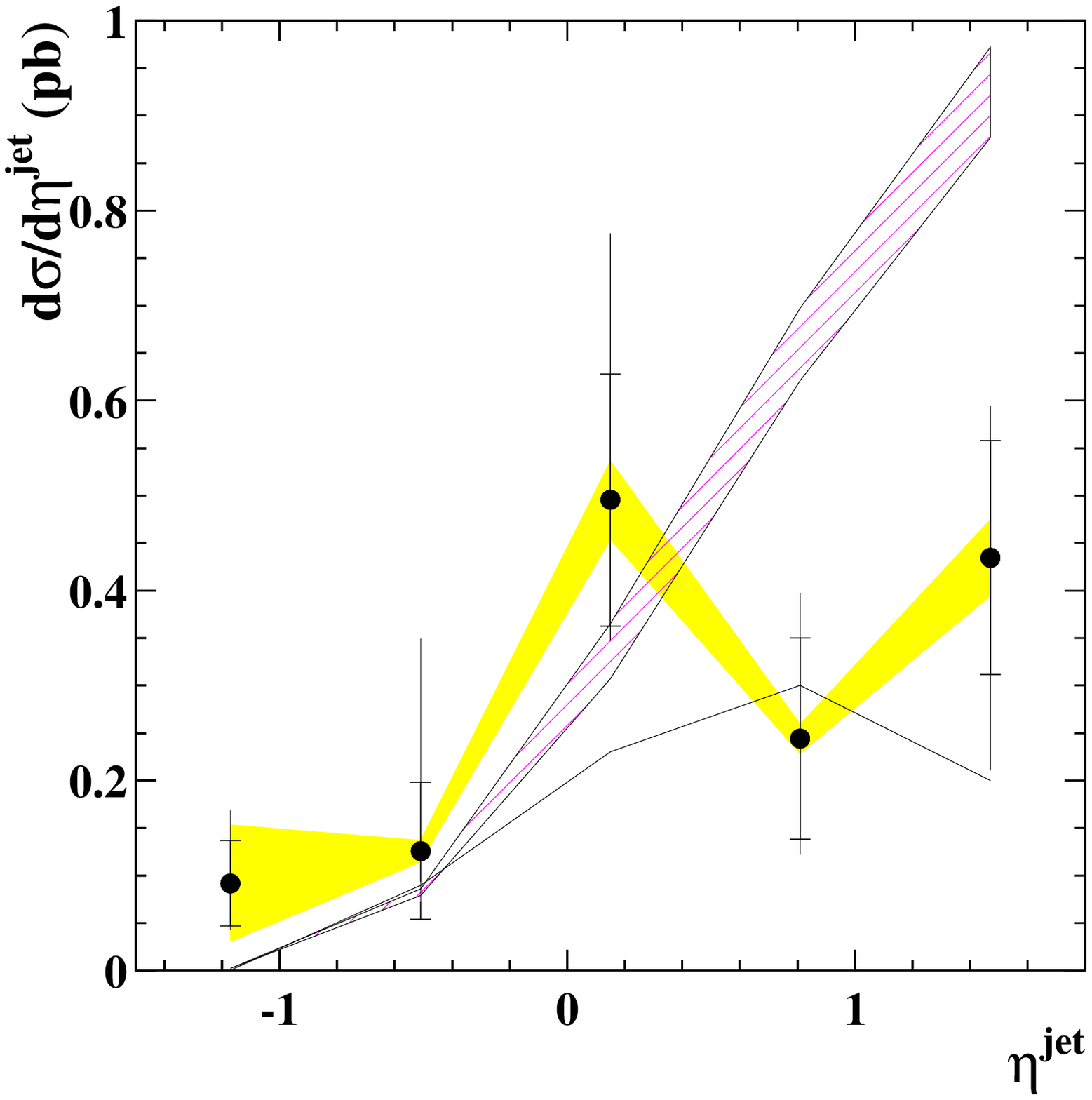,width=9cm}}
\put (9.0,0.6){\epsfig{figure=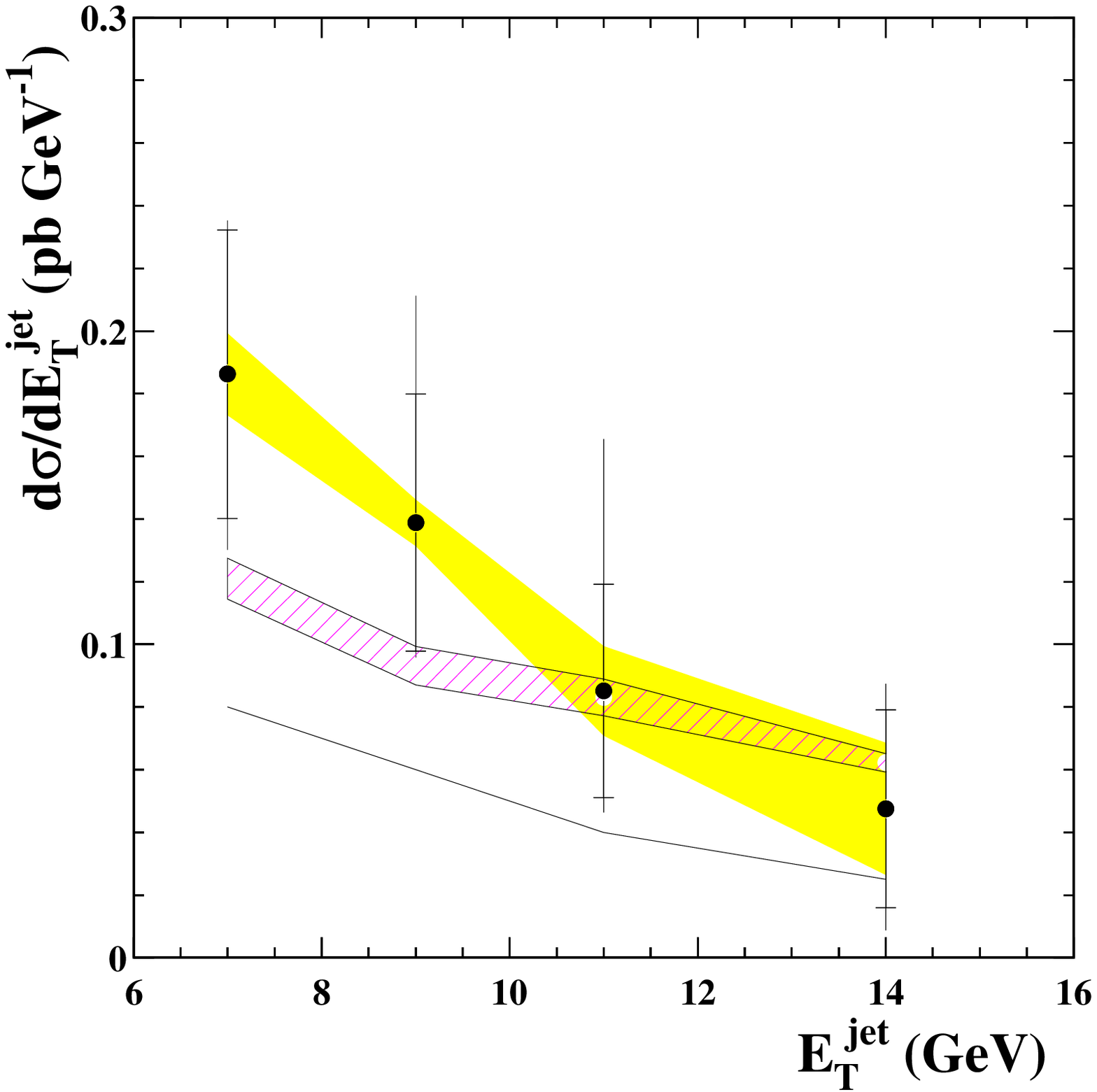,width=9cm}}
\put(0.9,18.5){\bf\Large\centerline{ZEUS}}
\put (7.3,17.4){\bf\small (a)}
\put (16.2,17.4){\bf\small (b)}
\put (7.3,8.0){\bf\small (c)}
\put (16.2,8.0){\bf\small (d)}
\end{picture}
\caption{\small Cross section for prompt-photon-plus-jet production
differential in
(a) photon rapidity,
 (b) photon transverse energy,
(c) jet pseudorapidity,
 (d) jet transverse energy,
 for events with a photon in the range
$-0.7 < \eta^\gamma < 0.9$ and $5 < E_T^\gamma < 10 {\rm\gev}$ and
one jet in the range $-1.5 < \eta^{\rm{jet}} < 1.8$ and
$E_T^{\rm{jet}} > 6 {\rm\gev}$. The inner error bars are statistical 
while the outer
represent systematic uncertainties added in quadrature.
The band around the data points
shows the effect of calorimeter energy scale uncertainty.
 The boxed band shows the parton-level
predictions of Kramer and Spiesberger 
 including
the effect of renormalisation scale uncertainty. The single line 
indicates
 their prediction of the contribution of photons radiated from the quark 
line.}

\label{fig5}
\vfill
\end{figure}

\end{document}